\documentclass[conference]{IEEEtran}
\IEEEoverridecommandlockouts

\usepackage{graphicx}
\usepackage{amsmath}
\usepackage[utf8]{inputenc}
\usepackage{color}
\usepackage{graphicx}
\usepackage{rotating}
\usepackage{tikz}
\usepackage{amssymb}
\usepackage{epsfig}
\usepackage{subfigure}
\usepackage{epstopdf}
\usepackage{algorithm}
\usepackage{algorithmic}
\usepackage{amsmath}
\usepackage{amsfonts}
\usepackage{dsfont}
\usepackage{url}
\usepackage{chemarr}
\usepackage{chemarrow}
\usepackage{bbold}
\usepackage[version=3]{mhchem}
\usepackage{mathabx}
\usepackage{tabularx,booktabs}
\usepackage{booktabs}

\def\BibTeX{{\rm B\kern-.05em{\sc i\kern-.025em b}\kern-.08em
    T\kern-.1667em\lower.7ex\hbox{E}\kern-.125emX}}
\begin{document}

\title{\huge{Communication in Plants: Comparison of  Multiple Action Potential and Mechanosensitive Signals with Experiments} }

\author{Hamdan Awan, Kareem Zeid, Raviraj S. Adve, Nigel Wallbridge,\\ Carrol Plummer, and Andrew W. Eckford%
\thanks{Note: This work was presented in part at the IEEE ICC 2019.}%
\thanks{Hamdan Awan, Kareem Zeid and Andrew W. Eckford are with the Department
of Electrical Engineering and Computer Science, York University, Toronto, Ontario, Canada M3J 1P3. E-mails: hawan@eecs.yorku.ca, aeckford@yorku.ca.}%
\thanks{Raviraj S. Adve is with The Edward S. Rogers Sr. Department of Electrical and Computer Engineering, University of Toronto, Toronto, Ontario, Canada M5S 3G4. E-mail: rsadve@ece.utoronto.ca}%
\thanks{Nigel Wallbridge and Carrol Plummer are with Vivent SaRL, 1299 Crans-pr\`es-C\'eligny, Switzerland. E-mails: nigel.wallbridge@vivent.ch, carrol.plummer@vivent.ch}%
}

\maketitle

\begin{abstract}

Both action potentials and mechanosensitive signalling are an important communication mechanisms in plants. Considering an information-theoretic framework, this paper explores the effective range of multiple action potentials for a long chain of cells (i.e., up to 100) in different configurations, and introduces the study of multiple mechanosensitive activation signals (generated due to a mechanical stimulus) in plants. For both these signals, we find that the mutual information per cell and information propagation speed tends to increase up to a certain number of receiver cells. However, as the number of cells increase beyond 10 to 12, the mutual information per cell starts to decrease. To validate our model and results, we  include an experimental verification of the theoretical model, using a PhytlSigns biosignal amplifier, allowing us to measure the magnitude of the voltage associated with the multiple AP's and mechanosensitive activation signals induced by different stimulus in plants. Experimental data is used to calculate the mutual information and information propagation speed, which is compared with corresponding numerical results. Since these signals are used for a variety of important tasks within the plant, understanding them may lead to new bioengineering methods for plants.

\end{abstract}

\IEEEpeerreviewmaketitle

\section{Introduction}
\label{sec:intro}

Plants need to respond to a variety of external stimuli in order to survive \cite{huber2016long}. Many different types of signals exist which transfer information concerning these stimuli from one cell to another. Plants use this information to regulate various important functions, such as growth or defense in response to their external environment. For example, plants sense light levels and use this information to alter the production of energy using photosynthesis \cite{surova2016variation}. Two important types of information signals in plants are electro-chemical signals, known as action potential (AP) signals \cite{sukhov2009mathematical,awan2019communication} and mechanosensitive signals \cite{sukharev1994large}. The main aim of these signals is to convey the information to neighbouring cells in the plant so that they can respond to an external stimulus by taking appropriate action. 

\textcolor{black}{From the literature we learn that there are two differences between both these types of signals. First, the type of stimulus generating them i.e., the AP signals are generated as a result of an external stimulus such as change in temperature or intensity of light, whereas the  mechanosensitive signals are generated as a result of mechanical stimuli such as touch or wind \cite{awan2,sachs1998mechanosensitive}. Second, these signals are usually generated in different types of plants, therefore these signals do not mix or interfere with each other. Furthermore the waveform of both these signals is similar as shown in \cite{awan2019information}. } Both types of signals propagate from transmitter to receiver cells by either diffusion or fast (active) movement of molecules from one cell to another in a group of cells. Since both these signals are generated by an external stimulus, it is important to study them using experiments.

We have learned from the models in \cite{evans2017chemical,vodeneev2018parameters,sukhova2017mathematical} that some signals in plants are informed by the phenomena of molecular communication \cite{nakano2013-book,farsad2016comprehensive}. The paradigm of molecular communication is inspired by the communication among living cells as suggested in \cite{Akyildiz:2008vt,awan2017improving,awan2019molecular,awan2016generalized1,awan2017molecular}. In previous works \cite{awan2019communication,8647210,awan1,awan2019impact} we have used molecular communication theory to study the communication properties of different signals in plants with different configurations of receiver cells. A key characteristic of molecular communication is the use of molecules to transmit information (in AP and mechanosensitive signals) from the transmitter to the receivers. The transmission of signalling molecules can be carried out by different mechanisms such as diffusion \cite{Pierobon:2010kz,awan2016demodulation,riaz2018using} or active transport \cite{farsad2011simple}. The information carrying molecules propagate to the receiver where the molecular communication system uses reactions such as linearized ligand-receptor binding, to decode the transmitted information \cite{Chou:2014jca}. In this work we use molecular communication theory to study the communication effectiveness of multiple AP and multiple mechanosensitive activation signals.

In previous works, we have presented the mathematical models for the generation of single and multiple AP signals in plants with up to five receiver cells, which may propagate by diffusion or active movement of molecules  \cite{awan2019communication,awan1}. In this paper we extend this study to include the generation of multiple AP's in a system with a higher number of cells (up to 100) as compared to only five in previous works. The main aim of this study is to understand the effective range of communication, in terms of the number of receiver cells, for these signals. 

In a recent paper \cite{awan2} we derived a model for the generation of single mechanosensitive activation signals in plants due to a mechanical stimulus.  In this paper we extend this work by presenting a mathematical model for the generation of multiple mechanosensitive activation signals. We also study the effective range of communication, in terms of number of receiver cells, of these signals. 
For both the multiple AP and multiple mechanosensitive signals we study the mutual information and information propagation speed for an increasing number of receiver cells, up to 100. This will help us understand the communication performance of these signals for different configurations of receiver cells in a system.

In this work we perform experiments on different plants to measure the amplitude and duration of voltage changes associated with multiple AP and multiple mechanosensitive activation signals. The purpose of the experimental work is to verify the results obtained by the theoretical models presented in this paper and in previous works. We use a PhytlSigns device developed by Vivent SaRL to carry out the experimentation and compare the magnitude and duration of the signals with the corresponding numerical results. We also use the experimental data to compute the mutual information and information propagation speed and compare these results with the corresponding theoretical values.  This experimental verification helps justify the use of information theoretic modelling to study biological communications in higher organisms such as plants.
 
Specifically, this paper makes the following three contributions. First, an extended mathematical model for generation of multiple mechanosensitive activation signals (due to mechanical stimulus) in plants is presented. Second, the communication effectiveness of systems comprising either multiple AP or multiple mechanosensitive activation signals are studied using mutual information and information propagation speed in systems with a higher number of receiver cells. Third, the theoretical models of the communication properties of both multiple AP and multiple mechanosensitive activation signals are verified using experimental measurements of voltage variations. 

\section{Related Work}
 
In this paper we consider two of the common types of signals which exist in plants, namely AP signals and mechanosensitive signals. AP signals are one of the many different types of electro-chemical signals existing in nature. In literature the role of electro-chemical signals and the associated physiological or biochemical response is studied in \cite{fromm2007electrical,fromm1995biochemical}.  In \cite{gilroy2016ros} the authors have studied the transmission  mechanism of electro-chemical signals from transmitter cells to the receiver cells. In \cite{sukhov2016electrical,szechynska2017electrical} the authors have suggested that electro-chemical signals can influence different processes of plants such as photosynthesis.  The different roles of electrical signals in plant physiology are analyzed in  \cite{surova2016variation,sukhov2015variation,sukhov2017high}. In some other works the authors have studied the role of electro-chemical signals in various functionalities of plants such as respiration \cite{lautner2014involvement}, gene expression \cite{pena1995signals}, hormone production \cite{hlavavckova2006electrical}, ATP content, and others \cite{surova2016variation}. 

Action potential electro-chemical signals (APs) are  generated in different plants such as {\em Venus flytrap} and {\em Mimosa pudica} in the presence of an external stimulus \cite{sukhov2011simulation}. Different mathematical models for the generation of electro-chemical AP signals and their influence on the physiological activity of plants are presented in \cite{sukhova2018influence,sherstneva2015participation}.  We can define an AP signal as a sudden change or increase in the resting potential of the cell as a result of some external \textcolor{black}{stimulus} \cite{sukhov2009mathematical}.  In a plant an AP signal propagates (in the form of molecules) from the transmitter to the receiver cells which are connected through plasmodesmata (a narrow thread of cytoplasm that allows communication between cells) \cite{fromm2007electrical}. In general an AP signal is produced as a result of passive fluxes of ion channels such as sodium, calcium and potassium channels \cite{fromm2007electrical,fromm1995biochemical,felle2007systemic}. Some models for AP generation in different plants are presented in \cite{sukhov2009mathematical,sukhov2011simulation}.  In our previous works we have presented a mathematical model for the generation of single and multiple AP signals in plants \cite{awan2019communication,awan1}. Previous work are limited to only a small number of receiver cells (5). Increasing the number of receiver cells enables us to draw insight into the effective range of AP signals in plants. In this paper we consider up to 100 receiver cells.

On the other hand mechanosensitive activation signals are generated in mechanosensitive ion channels in response to a mechanical stimulus such as touch or wind in plants. Initially discovered in living organisms in 1985 \cite{guharay1985mechanotransducer}, a significant amount of research has since been published on the topic of mechanosensitive signals in organisms such as bacteria, plants, animals and humans \cite{zhang2000mechanically,honore2006desensitization,sukharev1994large,cox2013selectivity}. Mechanosensitivity can simply be defined as a response of the living cells to a mechanical stimulus.   In a mechanosensitive organism, the mechanical stimuli modulates different physiological processes at the cellular level \cite{sachs1998mechanosensitive}. Mechanosensitive signals can induce different biochemical processes inside a cell that may be transient or long term in nature \cite{awan2}. From \cite{sachs1998mechanosensitive,valle2012mechanosensory} we learn that many organisms are responsive to a mechanical stimulus which is often in the form of stress, therefore the mechanosensitive ion channels are also known as stress gated ion channels. 

Mechanosensitive activation signals are generated in response to a mechanical stress which allows ions to flow across the cell membrane \cite{guharay1985mechanotransducer}. The biophysics of mechanosensitive signal generation is studied in many different works, such as in \cite{valle2012mechanosensory}, it is noted that mechanosensitive activation signals are generated in response to  stretch, vibration and touch. In \cite{markin2004thermodynamics}, we learn that theses signals are generated by the activation of the gating mechanism by  the tension in the bi-lipid layer of the cell membrane. The opening of gates leads to the flow of ions inside cell membrane such as Na$^+$, K$^+$ and Ca$^{2+}$. This flow of ions generate the mechanosensitive activation signal in the transmitter cell. Note that this activation signal is different from an action potential as it is generated through a mechanosensitive channel in response to a mechanical stress (cf. our previous work in \cite{awan2019communication}). The mechanosensitive signals play an important role in different functions of cell such as regulating growth of cellular organisms \cite{ladoux2017mechanobiology}. In our previous work \cite{awan2} we presented a model for generation of single mechanosensitive signals in plants. In this paper we aim to extend this model by presenting a mathematical model for generation of multiple mechanosensitive signals in plants. We follow it up by studying the effective range of multiple mechanosensitive signals in a system with  100 receiver cells. Finally in this paper we aim to experimentally verify the corresponding communication theoretical model for both the multiple AP and mechanosensitive signals in various plants. 

This paper is organized as follows. We describe the system model with different receiver configurations in Section \ref{system1a}. In Section \ref{system} we summarise previous work and present mathematical models for generation of multiple AP and mechanosensitive activation signals. Next we present the molecular communication model for the transmitter and propagation medium in Section \ref{system2a}. In Section \ref{complete} we present the molecular communication mathematical model for diffusion and active reactions. The expressions for mutual information and information propagation speed are presented in Section \ref{mutual}. This is followed by the experimental setup description in Section \ref{experiment}. \textcolor{black}{In Section \ref{numerical} we present the results and discussion.} Finally Section \ref{conclusion} presents the conclusion.

\begin{figure}
 \begin{center}
\includegraphics[trim=0cm 0cm 0cm 0cm ,clip=true, width=1\columnwidth]{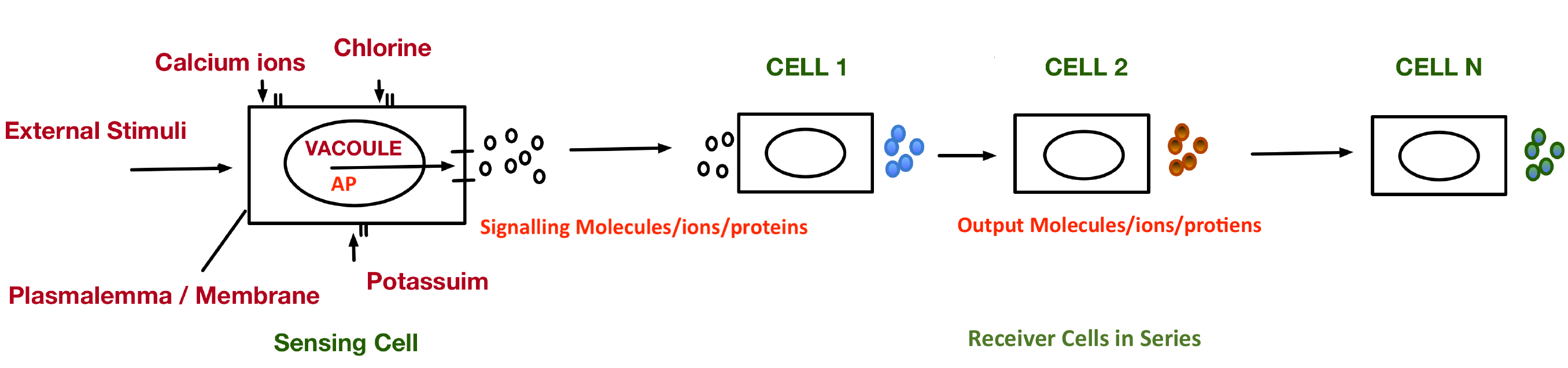}
 \caption{System Model-Series}
\label{system series}
 \end{center}
 \end{figure}
 
  \begin{figure}
 \begin{center}
\includegraphics[trim=0cm 0cm 0cm 0cm ,clip=true, width=1\columnwidth]{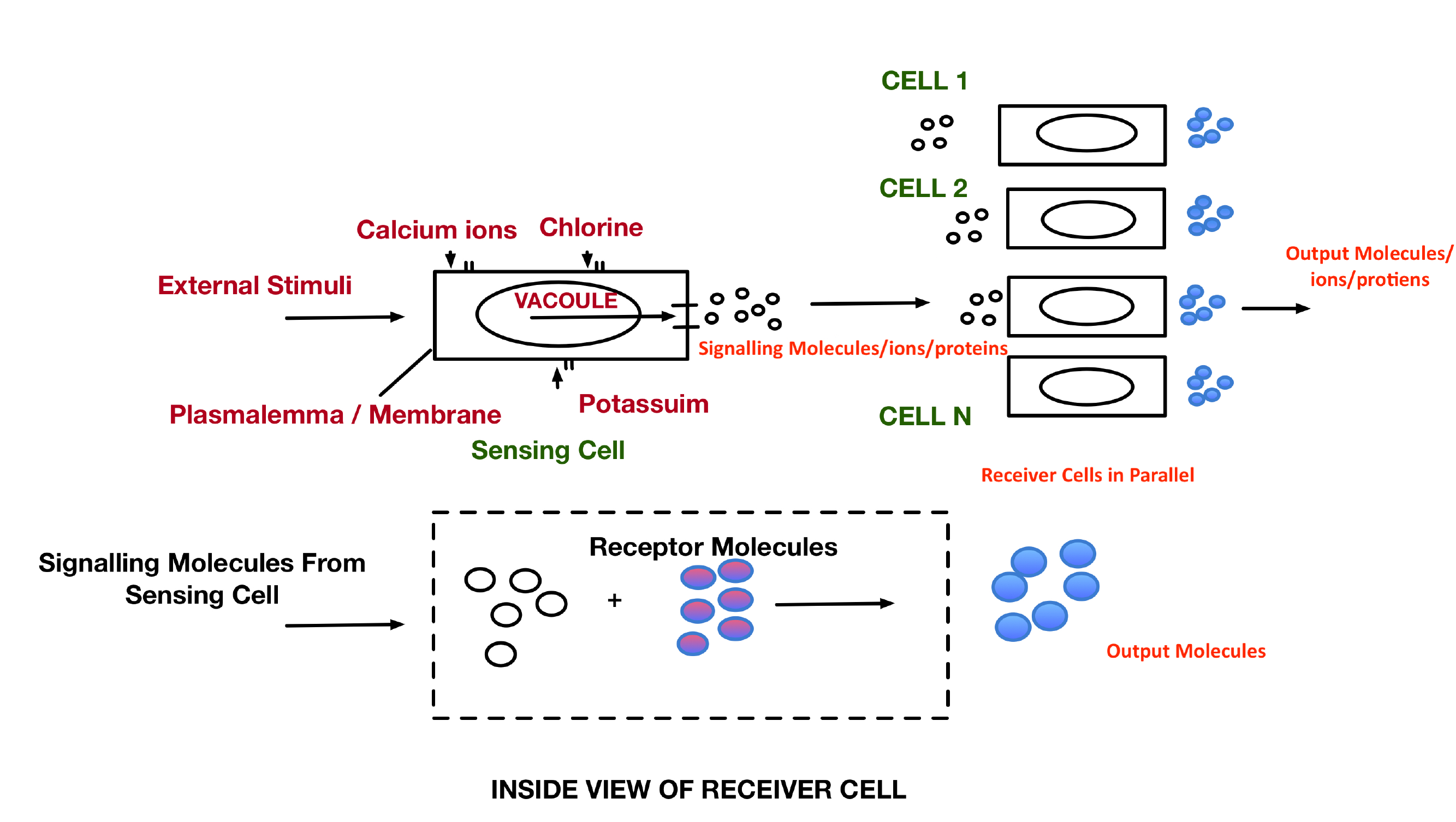}
 \caption{System Model-Parallel}
\label{system parallel}
 \end{center}
 \end{figure}

\section{System Model}
\label{system1a}
In this work we consider a system where different types of external stimulus will induce multiple AP and multiple mechanosensitive activation signals for inter-cellular communication in plants. We focus on studying the communication properties associated with these signals. The two types of external stimulus we use in this work are environmental (in form of temperature or light) and mechanical (in form of stress). To be specific we present two mathematical models i.e., (a) A model for the generation of multiple AP signals in electro-chemical systems due to an environmental stimulus. (b) A model for the generation of multiple mechanosensitive activation signals due to a mechanical stimulus.

\begin{figure}
        \centering
        \begin{subfigure}{\textwidth 1 cm}
            \includegraphics[trim=0cm 0cm 0cm 0cm ,clip=true, width=0.45\columnwidth]{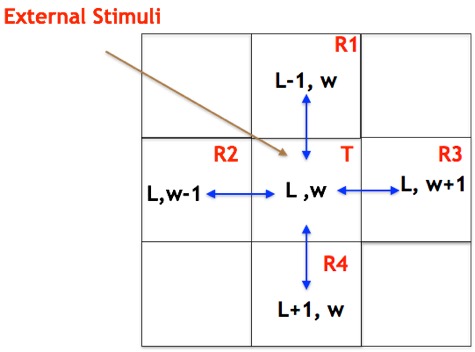}
        \end{subfigure}\hfill
            \begin{subfigure}{\linewidth 1 cm}
                \includegraphics[trim=0cm 0cm 0cm 0cm ,clip=true, width=0.45\columnwidth]{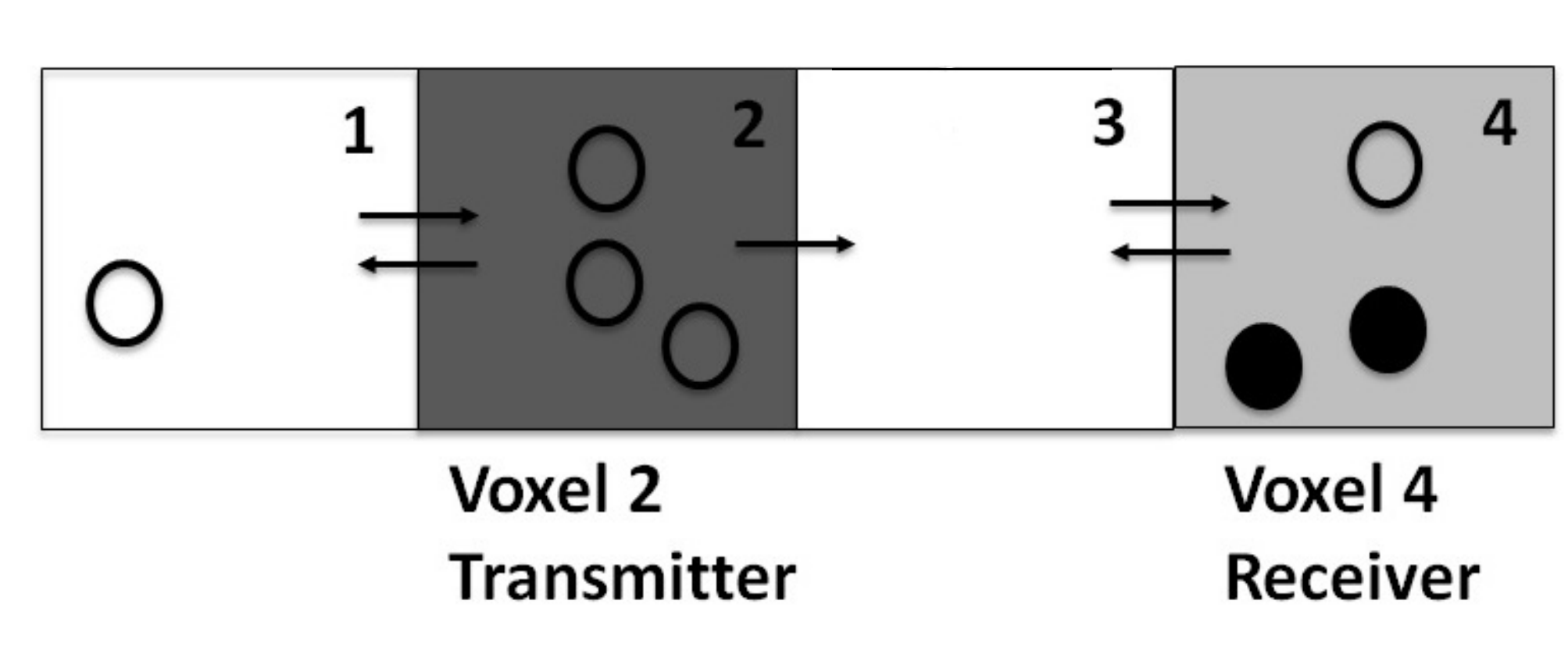}
        \end{subfigure}
            \caption{Voxel Model of Propagation-Parallel and Series}
            \label{1c}
    \end{figure}

In this work we use three different settings of receiver cells in system i.e, series, parallel (as shown in Figures \ref{system series} and \ref{system parallel}) and mixed configuration. Like our previous works we use a voxel model to describe the propagation of molecules between different cells as shown in Figure 3 for different configurations of cells. Note that for the sake of simplicity we have shown a 2-dimensional voxel setting in Figure 3 however, the analysis will be 3-dimensional. 

For both configurations of receiver cells we assume that an external stimulus generates multiple AP or multiple mechanosensitive activation signals in the transmitter cell.  The transmitter then releases signalling molecules depending on the intensity of the multiple AP or mechanosensitive activation signals. These signalling molecules act as the input to the receiver cells. The molecules can propagate from the transmitter to the receiver cells using different mechanisms i.e., diffusion based propagation and fast active propagation of molecules.  As the signalling molecules reach the receiver, they react with receptors to produce output molecules. The number of these output molecules in the receiver over time is the output signal of the molecular communication link. We aim to compute the mutual information between the input and output for the cases of multiple AP and multiple activation signals. The mathematical models describing the generation of both these signals in plants is shown below.

\section{Signal Generation Models}
\label{system}

\subsection{Model For Multiple AP Signals} 
\label{system1}

In this section we briefly present the mathematical model for the generation of multiple AP signals that is presented in our previous work \cite{awan1}. For $E_R$, the resting potential of the transmitter cell, the expression for the new membrane potential $E_m$ as a result of an external stimulus (causing a change in ion-concentrations) is given as:
\begin{equation}
E_m= \frac{g_k E_k + g_{cl} E_{cl} + g_{ca} E_{ca}}{g_k + g_{cl}+ g_{ca}}
\label{1:sa}
\end{equation}
 \begin{equation}
\quad \textrm{where,} \quad  g_i = \frac{Fh_i}{E_R-E_i}
\end{equation}
where the general term $g_i$ represents the electrical conductivity of an ion channel $i$, $E_i$ represents the resting potential value for the ion channel $i$, $F$ represents Faraday's constant and \textcolor{black}{$h_i$} is the ion flow across the membrane which is given as follows:
\begin{equation}
h_i = z  \mu P_m p_o \frac{\phi _i \eta_o - \phi_o \eta_i (\exp (-z \mu))}{1-\exp (-z \mu)} 
\end{equation}

The term $\mu$ denotes the normalized resting potential of cell membrane and is given as:
\begin{equation}
\mu = \frac{E_m F}{R_c T_c}
\end{equation}

The values of these parameters are presented in Table \ref{table:1}. To explain the terms in Eqs. (3) and (4): $z$ is the ion charge; $R_c$ is the gas constant; $T_c$ represents the temperature; $\phi_i$ (resp. $\eta_i$) is the probability that the ion is (resp. is not) linked to the channel on the inside; $\phi_o$ (resp. $\eta_o$) is the probability that the ion is (resp. is not) linked to the channel on the outside; $P_m$ represents the \textcolor{black}{maximum permeability of the cell}; and $p_o$ represents the ion-channel opening state probability as a result of change in the concentrations of ions. For $k_1$ (opening)  and $k_2$ (closing) reaction rate constants we obtain $p_o$ as:
\begin{equation}
\frac{dp_o}{dt} = k_1 (1- p_o) - k_2 (p_o)
\end{equation}
\textcolor{black}{ Note that the channel $k_1$ (opening)  and $k_2$ (closing) reaction rate constants depend  on the membrane potential crossing a specific threshold value, resulting in the generation of an AP signal. We also note that these rate constants are exponentially dependent on the membrane potential as discussed in \cite{sukhova2017mathematical}. }

\begin{figure}
\begin{center}
\includegraphics[trim=0cm 0cm 0cm 0cm ,clip=true, width=1\columnwidth]{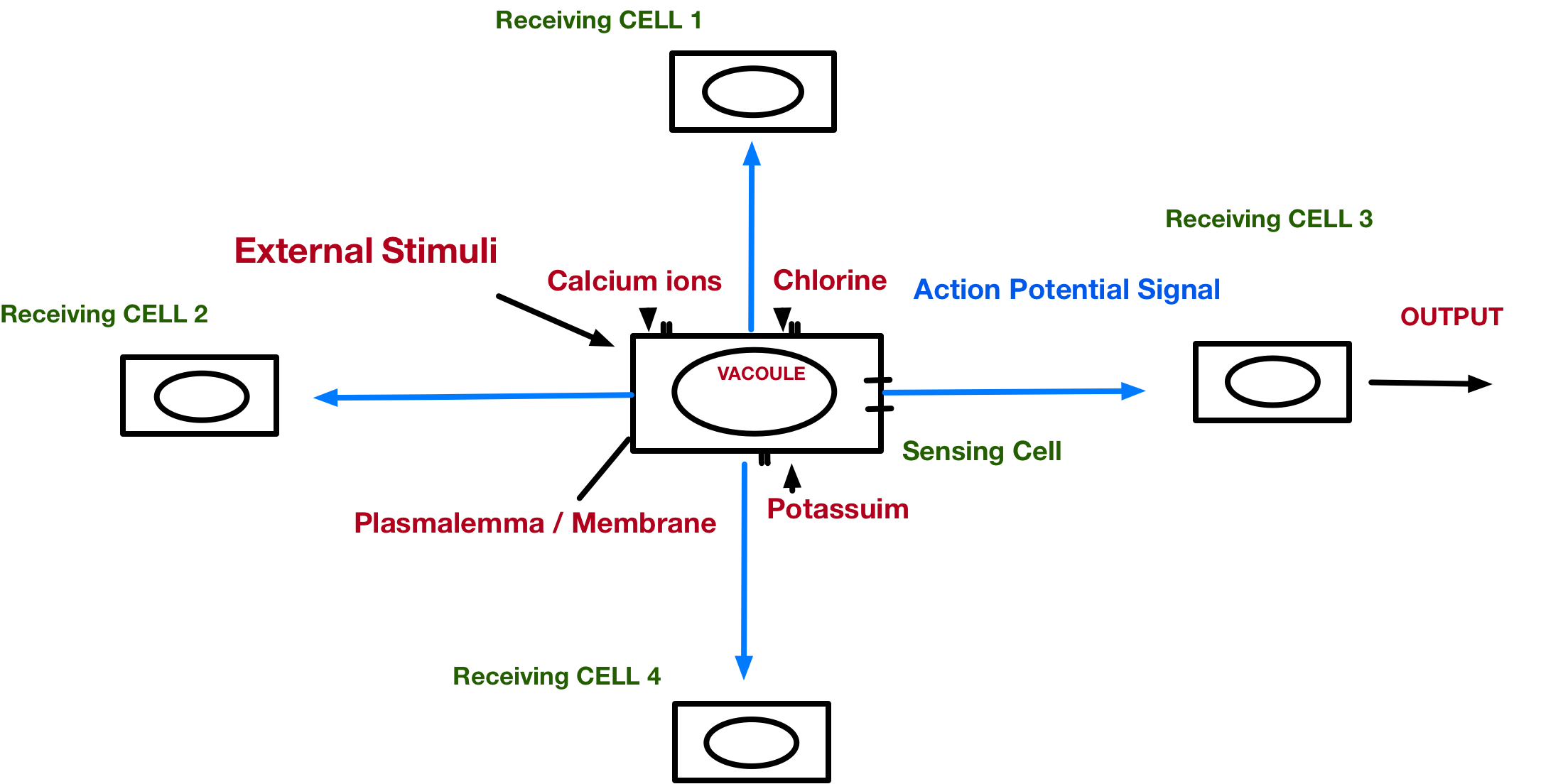}
\caption{Multiple AP Model }
\label{1cD}
\end{center}
\end{figure}

The AP signal propagates from one cell to another, resulting in multiple APs impacting a single cell. We use a lattice of cells in the form of voxels to model the impact of multiple APs. In this paper we consider a 3-d model of cells where there is a limit of about 12 cells that can be in contact with a transmitter cell as the cells are considered to be closely packed. We also assume that strong electrical coupling exists between the cells. For the ease of understanding we show an example of 2-d lattice of cells with dimensions length ($L$) $\times$ width ($w$) in Figure 4. The transmitter cell $T$ in the centre is surrounded by 4 receiver cells $R_1$-$R_4$ on each side and multiple AP signals can propagate to or from each of these cells.  The number of molecules flowing as a result of multiple AP signals depends on the inter-cellular interactions and is given by:
\begin{equation}
n_{Lw} = \sum_{K,s} G_{LwKs}  (E_m^{Ks}- E_m^{Lw})
\end{equation}

Where $L$ = $[1,L_T]$ and $w$ = $[1,w_T]$. $L_T$ and $w_T$ represent the total length and width of the mesh of voxels. $E_m^{Ks}$ and $E_m^{Lw}$ represent the cell membrane potentials at the transmitter cell location $(L;w)$ and general receiver cell location $(K;s)$ respectively. The term $ G_{LwKs}$ represents the electrical conductivity between the transmitting cell $(L;w)$ and the general receiver cell $(K;s)$.  Note that Eq. \eqref{1:sa} only accounts for the change in the membrane potential due to a single AP. However the change in the potential can also occur as a result of multiple Action potential signals from neighboring cells. This leads to the following equation:
 \begin{equation}
 E_m^{Lw}= \frac{g_k E_k + g_{cl} E_{cl} + g_{ca} E_{ca} + \sum_{K,s} G_{LwKs}  E_m^{Ks} }{g_k + g_{cl}+ g_{ca} + \sum_{K,s} G_{LwKs}}
 \label{1cx}
\end{equation}
Therefore the change in the potential can happen for two reasons i.e. (a) one or more AP signals from the neighbouring cells which introduces the additional term $G_{LwKs}$ i.e. the electrical conductivity between the sensing cell $T$ and the general neighbor cell $R_i$.
and (b) an external stimulus generating an AP signal within the plant cell which propagates to receiver cells.  \textcolor{black}{The input of the system $U(t)$ i.e. the number of signalling molecules emitted by the transmitter cell driven by the multiple AP is dependent on $E_m^{Lw}$:}
\begin{equation}
U(t) \; \propto \; E_m^{Lw}
\label{1:ua}
\end{equation}\textcolor{black}{Where $E_m^{Lw}$ (which depends on the channel opening probability $p_o$ in Eq. (5) and other parameters in Eqs. (3) and (4) presented in Table II) is given by Eq. \eqref{1cx}. Note that this $U(t)$ acts as the system input. This relation means that when multiple AP signals are generated the transmitter emits a higher number of molecules as compared to a single AP signal. }

\subsection{Model For Multiple Mechanosensitive Activation Signals} 
\label{system2}
 In this section we present the model for generation of multiple mechanosensitive activation signals in response to a mechanical stimulus. In plants, the mechanical sensitivity of ion channels is similar to that of voltage sensitivity due to an environmental stimulus. For mechnosensitive channels the effect of stress or tension is to open the ion channels resulting in the flow of ions across the cell membrane. A mechanosensitive channel can be defined as a stress-dependent equilibrium between the open and closed states as described in \cite{sukharev1997mechanosensitive}:
 \begin{align}
C \underset{}{\stackrel{K'}{\rightleftharpoons}} O	  
\end{align}
where $K'$ is the equilibrium constant between the closed and open states. If the cross-sectional area of the ion channel has a non-zero difference (between the two states) $\Delta A$ = $A_{open}$- $ A_{closed}$, then the contribution of membrane tension $\sigma$ in the free energy of channel opening,  $\Delta G$ is given by \cite{haswell2011mechanosensitive}:
\begin{align}
\Delta G = \Delta G' -  \sigma	\Delta A  
\end{align}
where $\Delta G'$=$- R_c T_c\ln K'$ is the free energy associated with the opening of channel in the absence of stress. $R_c$ is the gas constant and $T_c$ is the absolute temperature. The expression for the stress $\sigma_{1/2}$ required to open half the ions channels at equilibrium is obtained as follows:
\begin{align}
\sigma_{1/2} = \frac{\Delta G}{\Delta A  }	
\end{align}
From this expression we learn that the channel response in the presence of stress is highly dependent on the cross-sectional area. This expression is also analogous to the trans-membrane voltage drop $V_{1/2}$ when half the channels are open in the presence of charge $z$ moving across the cellular membrane. The sensitivity of the conformational equilibrium to the stress is encoded by:
 \begin{align}
\Delta A = \frac{d \Delta G}{d  \sigma }	
\end{align}
which shows that as $\Delta A$ increases, the shift in the equilibrium increases in response to an applied stress. Therefore the channel is said to be more mechanosensitive. Assuming that the channel complies with the stress which results in $ \Delta A$ expansion of the membrane, the work produced by the external force is given as:
\begin{align}
W_m = \sigma \Delta A	
\end{align}
This suggests that as $\Delta A$ increases, it results in an increased flow of ions leading to a higher amplitude of activation signal. For a mechanosensitive ion channel the opening probability depends on different factors such as the applied stress, $\Delta G$ and $\Delta A$, and is obtained by using Boltzmann relation \cite{sukharev1997mechanosensitive}:
 \begin{align}
\frac{P_o}{1 - P_o }= \exp [{(\sigma  \Delta A- \Delta  G)}/K_B \sigma ]
\end{align}
where $K_B$ is the Boltzmann constant. Using some mathematical work we  reduce this equation to the following simplified form similar to \cite{sachs2010stretch}.
 \begin{align}
P_o= \frac{1}{1 + e^{-\Delta  G/K_B \sigma}}
\end{align}
As a result of the opening of mechanosensitive ion channels an activation signal is generated. We can obtain the voltage corresponding to this signal as:
\begin{equation}
V= \frac{g_k V_k + g_{Na} V_{Na} + g_{ca} E_{ca}}{g_k + g_{Na}+ g_{ca}}
\label{vs}
\end{equation}
where $g_i$ is the general conductance of a mechanosensitive ion channel $i$. This conductance can be defined as the proportionality between the voltage drop across the cell membrane and the ionic current flowing through the channel. The general term $V_i$ is the non-zero initial voltage corresponding to the activation signal in channel $i$. The term $g_i$ is given as:
 \begin{equation}
\quad \textrm{where,} \quad  g_i = \frac{Fh_i}{V_i}
\end{equation}
where $F$ is Faraday's constant and \textcolor{black}{$h_i$} is the ion $i$ flow across the membrane and is given as:
\begin{equation}
h_i = z  \mu P_m P_o \frac{\phi _i \eta_o - \phi_o \eta_i (\exp (-z \mu))}{1-\exp (-z \mu)} 
\end{equation}
where the term $\mu$ denotes the normalized initial potential corresponding to the activation signal. The rest of the terms are explained in Tables \ref{table:1} and III.  Note that $h_i$ depends on the channel opening probability $P_o$ of the mechanosensitive ion channel $i$ which is obtained from Eq. (15).

The expression in Eq. (16) only accounts for the change in the membrane potential due to a single mechanosensitive activation signal. However the change in potential can also occur as a result of multiple activation signals flowing from neighboring cells. For example we again consider the example in Figure 4 where the transmitter cell $T$ in the centre is surrounded by 4 receiver cells $R_1$-$R_4$ on each side. Multiple activation signals can travel to or from each of these cells to the transmitter cell. For $L$ = $[1,L_T]$ and $w$ = $[1,w_T]$, where $L_T$ and $w_T$ represent the total length and width of the mesh of voxels, the number of molecules flowing as a result of multiple activation signals depends on the inter-cellular interactions and is given by

\begin{equation}
n_{Lw} = \sum_{K,s} G_{LwKs}  (V^{Ks}- V^{Lw})
\end{equation}

Where $V^{Ks}$ and $V^{Lw}$ represent the cell membrane potentials at position $(L;w)$ and $(K;s)$ respectively. The term $ G_{LwKs}$ represents the electrical conductivity between the sensing cell $(L;w)$ and the general neighbor cell $(K;s)$.

This leads to the following equation:
 \begin{equation}
 V^{Lw}= \frac{g_k V_k + g_{Na} V_{Na} + g_{ca} V_{ca} + \sum_{K,s} G_{LwKs}  V^{Ks} }{g_k + g_{Na}+ g_{ca} + \sum_{K,s} G_{LwKs}}
 \label{1cv}
\end{equation}

where the term $G_{LwKs}$ represents the electrical conductivity between the sensing cell $T$ and the general neighboring cell $R_i$. Therefore the change in potential can occur in two ways  i.e. (a) one or more AP signals from the neighbouring cells and (b) an external stimulus generating an AP signal within the transmitter cell which propagates to receiver cells. Using $V^{Lw}$ from Equation \eqref{1cv} we can obtain the current flowing $I^{Lw}$ through the mechanosensitive channel by using ohm's law. The input of the system $U(t)$ i.e., the number of molecules emitted by the transmitter cell is given by:
\begin{equation}
U(t) \; \propto \; V^{Lw} \propto \; I^{Lw}
\label{1:ua2}
\end{equation}
where $V^{Lw}$ (depending on the channel opening probability $p_o$ and other parameters in Tables II and III) is given by Eq. \eqref{1cv}. This relation means that when the magnitude of activation signal voltage increases in the presence of a mechanical stimulus, the transmitter emits an increased number of molecules.

\section{Transmitter and Propagation Model}
\label{system2a}

\subsection{Transmitter Cell}

The overall system model considered in this paper for different configurations of receiver cells is shown in Figures \ref{system series} and \ref{system parallel}. We also consider a mixed (hybrid) configuration of receiver cells in this work. For each case we have a single transmitter cell which releases signalling molecules in the presence of an external stimulus. The number of molecules released depends on the magnitude of multiple AP or multiple mechanosensitive activation signals. These signalling molecules carry the information to the receiver cells. In the next subsection we explain the propagation mechanism of signalling molecules by using a voxel setting.

\subsection{\textcolor{black}{Voxel Model for Propagation}}
For the propagation of molecules to neighboring cells we assume that the cells in the system are closely packed. This means that molecules exiting from the membrane of one cell can enter the membrane of the neighboring cell(s) either through random diffusion or fast active propagation. To model this we assume a voxel setting for the system as shown in Figure 3 for both the parallel and series configuration. We divide the overall medium into $M_x \times M_y \times M_z$ cubic voxels where the volume of each voxel is $ \Delta^3$ representing a single cell. We assume the medium as a three dimensional space of dimension $\ell_X \times \ell_Y \times \ell_Z$ where each dimension is an integral multiple of length $ \Delta$ i.e. $\ell_X = M_x \Delta$, $\ell_Y = M_y\Delta$ and $\ell_Z = M_z \Delta $. In Figure \ref{1c} we show two examples i.e. parallel case with $M_x = M_y =3$ and $M_z = 1$ and series case with $M_x = 4$ and $M_y = M_z$ = 1. For the parallel case the transmitter and receiver are located at the voxels with indices $T$ and $R_i$. The different arrows in Figure 3 indicate the direction of propagation of molecules in the voxels. 

\section{Diffusion-Reaction Mathematical Model}
\label{complete}
 The overall state of the system can be defined as the number of molecules in each voxel of a 3-dimensional medium. Let $n_{L,i}$ denote the number of input signaling molecules in a general voxel $i$. The number of signalling molecules released by the transmitter depends on the amplitude of multiple AP or multiple activation signals generated as a result of an external stimulus. Assuming that the transmitting cell and each receiver cell occupy one voxel, we define the state of system for the series configuration of cells shown in Figure 3 (i.e., $i$ = 4 voxels in series) as:
  \begin{equation}
n_L (t) = [n_{L,1}(t),n_{L,2}(t),n_{L,3}(t) ,n_{L,4}(t)]^T
\end{equation}
where the superscript $T$ denotes the matrix transpose. We consider two propagation mechanisms of molecules between the transmitter and receiver cells i.e., (a)  Diffusion of molecules (b) Fast (active) propagation. This means that we can model the system as either a diffusion-reaction system or a reaction only system.  Using the approach presented in \cite{awan1} we derive the stochastic differential equation (SDE) governing the dynamics of the system. The SDE equation for a diffusion only system with a diffusion matrix $H$ is given as:

\begin{align}
\dot{n}_L(t) & = H n_L(t)  + \sum_{j = 1}^{J_d} q_{d,j} \sqrt{W_{d,j}(n_L(t))} \gamma_j 
+ {\mathds 1}_T U(t)
\label{eqn:sde:do} 
\end{align}
where, for the series voxel setting in Figure 3, we have
\begin{align}
H =  
\left[ \begin{array}{ccccc}
-d & d & 0 & 0  \\
d & -2d & 0 & 0  \\
0 & d & -d & d   \\
0 & 0 & d & -d  \\
\end{array} \right] 
\label{eqn:H} 
\end{align}  
In Eq. \eqref{eqn:sde:do} the term $J_d$ refers to diffusion jump vector and $W_{d,j}(n_L(t))$ is the corresponding jump function. 

On the other hand the reaction only system includes the reactions of incoming signaling molecules $L$ with the receiver to produce the output molecules $X$. The number of output molecules produced over time is the output signal of the system. In this paper we use a simple example of a linearized ligand-binding type receiver which consists of following reactions (along with respective jump functions and jump rates).
\begin{align}
L &  \rightarrow X 		& \left[ \begin{array}{cc} -1 & 1 \end{array} \right]^T&, k_+ n_{L,R}  \label{cr:rc1}  \\
X & \rightarrow L		& \left[ \begin{array}{cc} 1 & -1 \end{array} \right]^T&, k_- n_X       \label{cr:rc2}
\end{align}
where $k_+$ and $k_-$ are the reaction rate constants. The term $n_{L,R}$ is the number of signaling molecules in the receiver voxel and $n_X$ is the number of output molecules in the receiver. Similar to the diffusion only case, the state vector and SDE for the reaction only system is given as:
\begin{align}
 \tilde{n}_R(t) 
 & =  \left[ \begin{array}{c|c}
 n_{L,R}(t) & n_X(t)  
\end{array} \right]^T 
\end{align} 
\begin{align}
\dot{\tilde{n}}_R(t) & = R \tilde{n}_R(t) + \sum_{j = 1}^{J_r} q_{r,j} \sqrt{W_{r,j}(\langle \tilde{n}_R(t) \rangle)} \gamma_j + {\mathds 1}_T U(t) 
\label{eqn:sde:ro11} 
\end{align}
where $q_{r,j}$ and  $W_{r,j}$ represent the jump vectors and jump rates for the reactions in the system. The term $\gamma_j$ represents continuous white noise which accounts for the noise due to the reactions in the system. Similar to the diffusion matrix $H$  we define a reaction matrix $R$ shown in Table \ref{table:1sa}) which depends on the reactions in the receiver. 

\textcolor{black}{To model the complete diffusion-reaction system, we assume $n(t)$ as the state of the complete system which accounts for both the diffusion and reaction events in the system. The vector $n(t)$ contains the number of each type of molecules in each voxel (or cell). The physical meaning of $n(t)$ is that it describes how the molecules are distributed over the medium at time $t$, and is given by:}
\begin{align}
n(t) = 
 & \left[ \begin{array}{c|c}
 n_{L}(t)^T & n_X(t)  
\end{array} \right]^T
\label{eqn:state} 
\end{align}
The general SDE for the complete system which accounts for both the diffusion and fast propagation is given as:
\begin{align}
\dot{n}(t) & = { A} n(t) + \sum_{i = 1}^{J} q_{j} \sqrt{W_{j}(n(t))} \gamma_j + {\mathds 1}_T U(t) 
\label{eqn:mas11sv}
\end{align}
When we consider the diffusion propagation, the matrix $A$ will be the combination of $H$ and $R$ matrices. Whereas when we consider the fast active propagation of molecules, the entries of matrix $A$ will only depend on matrix $R$. The corresponding jump vectors and jump rates vary accordingly in each case.
\begin{table}[]
\centering
\caption{$R$ Matrix for different receiver circuits}
\begin{tabular}{|c|c|}
\hline
\multicolumn{1}{|c|}{Receiver }	&	\multicolumn{1}{|c|}{R Matrix}	\\
\hline
Reactions 25 and 26&    $ \begin{bmatrix}  -k_{+} & k_{-} \\ k_{+} & -k_{-} \end{bmatrix}$
\\ \hline
\end{tabular}
\label{table:1sa}
\end{table}
Next we use the Laplace transform of Equation \eqref{eqn:mas11sv} to obtain the number of output molecules in the receiver:
\begin{align}
  N_X(s)  = 
 \underbrace{   {\mathds 1}_X (sI - { A})^{-1} {\mathds 1}_T }_{\Psi(s)}   U(s)
\label{eqn:mas11b}
\end{align}
For the detailed step by step derivation of these equations refer to our previous work \cite{awan2019communication}. Next we use this expression to study different communication properties of the system.

\section{Mutual Information and Information propagation Speed}
\label{mutual}

\subsection{Mutual Information}
In this section we will derive the mutual information expression for the complete system using the expressions in Eqs. \eqref{eqn:mas11sv} and \eqref{eqn:mas11b}. We will present a general derivation which holds for the cases when the input number of signalling molecules depend on either the magnitude of multiple AP or for multiple mechanosensitive activation signals.  

To derive the mutual information we use the input number of molecules i.e. $U(t)$ from Eq. \eqref{1:ua} or \eqref{1:ua2} depending on the signal we are considering. The number of output molecules is given by Eq. \eqref{eqn:mas11sv}. To calculate the mutual information we use the approach described in \cite{awan2019communication}. From \cite{Tostevin:2010bo} we learn that for two Gaussian distribution random processes $a(t)$ and $b(t)$, the mutual information $I_m(a,b)$ is: 
\begin{align}
I_m(a,b) &= \frac{-1}{4\pi} \int_{-\infty}^{\infty} \log \left( 1 - \frac{|\Phi_{ab}(\omega)|^2}{\Phi_{aa}(\omega) \Phi_{bb}(\omega)}  \right) d\omega
\label{eqn:MI0}
\end{align} 
where $\Phi_{aa}(\omega)$ (resp. $\Phi_{bb}(\omega)$) is the power spectral density of $a(t)$ ($b(t)$), and $\Phi_{ab}(\omega)$ is the cross spectral density of $a(t)$ and $b(t)$. To apply this result for our system, we need a result from \cite{warren2006exact} which computes the power spectral density of a system consisting only of chemical reactions with linear reaction rates. Assuming all the jump rates $W_j(n(t))$ in Eq. \eqref{eqn:mas11sv} are linear, we obtain the power spectral density of $n(t)$ using following:
\begin{align}
\dot{n}(t) & = A n(t) + \sum_{i = 1}^{J} q_{r,j} \sqrt{W_{r,j}(\langle n(\infty) \rangle)} \gamma_j + {\mathds 1}_T U(t) 
\label{eqn:complete2}
\end{align} 
where $\langle n(t) \rangle$ denotes the mean of $n(t)$. The result in Eq.~\eqref{eqn:complete2} models a linear time-invariant (LTI) stochastic system subject to Gaussian input and Gaussian noise. The power spectral density $\Phi_{X}(\omega)$ of the output signal $n_X(t)$ is given as: 
\begin{align}
\Phi_{{X}}(\omega) & =  |\Psi(\omega) |^2 \Phi_u(\omega) + \Phi_{\eta}(\omega)
\label{2331a} 
\end{align}
where $\Phi_u(\omega)$ is the power spectral density of $U(t)$, $\Phi_\eta$ denotes the stationary noise spectrum. $|\Psi(\omega)|^2$ is the channel gain with $\Psi(\omega) = \Psi(s)|_{s = i\omega}$ and is given by:
\begin{align}
\langle N_{X} (s) \rangle  & = {\mathds 1}_X \langle N(s)  \rangle = 
\underbrace{ {\mathds 1}_X  (sI - A)^{-1} {\mathds 1}_T }_{\Psi(s) }   U(s)
\label{21a}
\end{align}
where $\Psi(s)$ incorporates both the consumption of input signalling molecules and the interaction between the input and output molecules. $\Phi_{\eta}(\omega)$ is the stationary noise spectrum and is given by: 
\begin{align}
\Phi_{\eta}(\omega) & =   \sum_{j = 1}^{J_r} | {\mathds 1}_X (i \omega I - A)^{-1} q_{r,j} |^2 W_{r,j}(\langle n_{}(\infty) \rangle) 
\label{eqn:spec:noise2} 
\end{align} 
where $\langle n_{}(\infty) \rangle$ is the mean state of system at time $\infty$ due to a constant input. By using the standard results on the LTI system, the cross spectral density $\Psi_{xu}(\omega)$ is:
\begin{align}
|\Psi_{xu}(\omega)|^2 &= |\Psi(\omega) |^2 \Phi_u(\omega)^2 
\label{eqn:csd} 
\end{align} 
Next by substituting Eq.~\eqref{2331a} and Eq.~\eqref{eqn:csd} into the mutual information expression in Eq.~\eqref{eqn:MI0}, we obtain the  mutual information $I(n_{X},U)$ between $U(t)$ and $n_{X}(t)$ as:
\begin{align}
I(n_{X},U) = \frac{1}{2} \int \log \left( 1+\frac{ | \Psi(\omega) |^2}{\Phi_{\eta}(\omega)} \Phi_u(\omega) \right) d\omega
\label{eqn:mi1}
\end{align}
The maximum mutual information of the communication link can be determined by applying the water-filling solution to Eq. \eqref{eqn:mi1} subject to a power constraint on the input $U(t)$ \cite{gallager1968information}. 

\subsection{Information Propagation Speed}
\label{infops}
To calculate the information propagation speed we use the mutual information calculated, for an increasing number of receiver cells in different configurations, for both multiple AP and multiple mechanosensitive activation signals. To obtain the propagation speed we select a threshold value of the mutual information and calculate the time difference at which the mutual information curve for different number of receiver cells crosses this threshold value. We use the following relation to calculate the information propagation speed:

\begin{equation}
  V_i=  \frac{1}{\mathbf{E} [\Delta t_{i,i+1}]}
  \label{infop}
\end{equation}
where $\Delta t_{i,i+1}$ is the time difference at which the mutual information, for an increasing number of receiver cells, crosses the threshold value. Note that \textcolor{black}{ $\mathbf{E}$ is the expectation operator.}

\section{Experimental Setup}
\label{experiment}

In this paper we have performed experiments to verify the results from the theoretical models. Specifically, we calculate the experimental results for the voltage of a multiple AP signal (in \textit{Mimosa pudica}) and multiple activation signals (in \textit{Aloe Vera}). The experimental setup is shown in Figures 5-6.

\begin{figure}
\begin{center}
\includegraphics[trim=0cm 0cm 0cm 0cm ,clip=true, width=0.8\columnwidth]{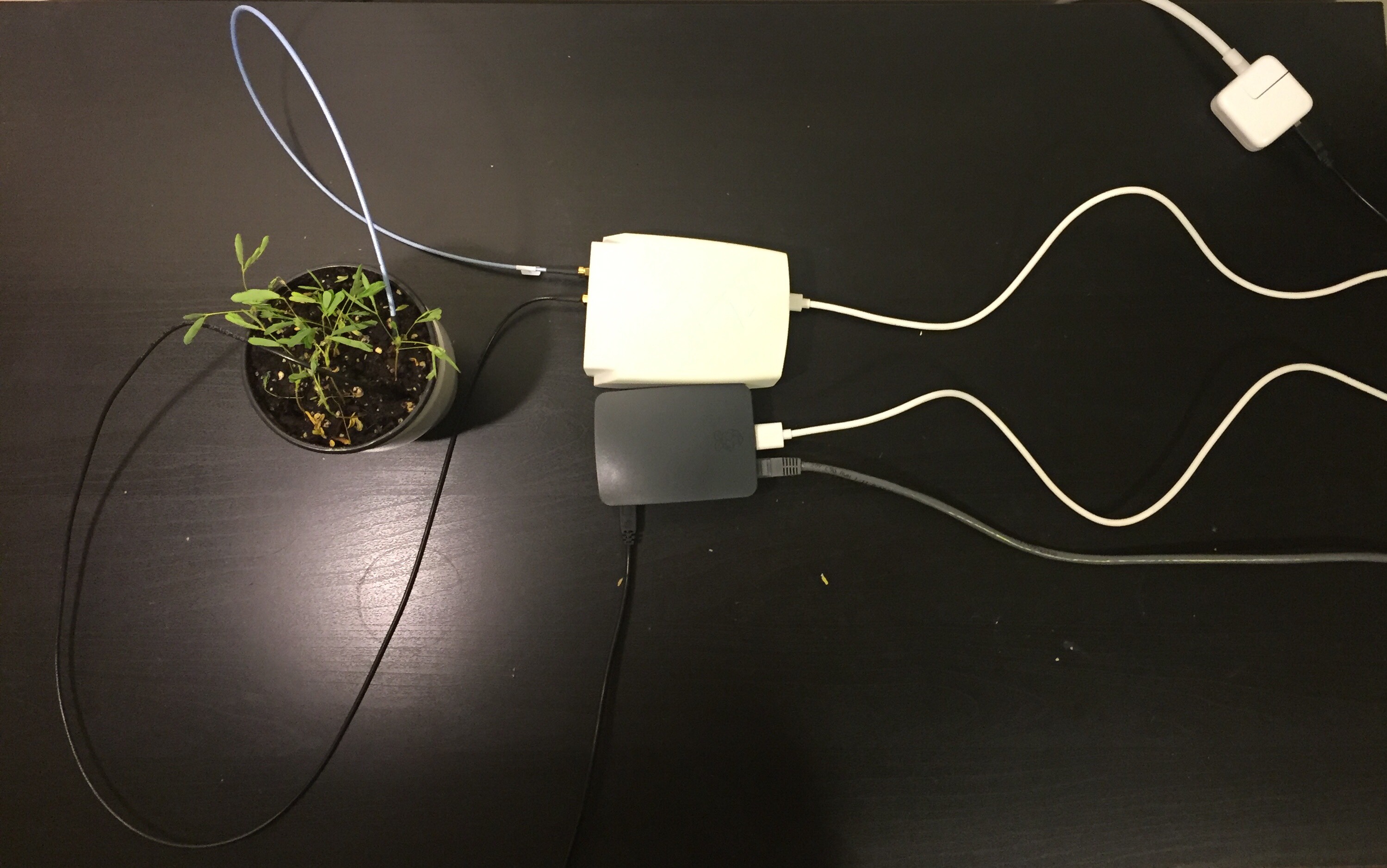}
\caption{Experimental Setup for Measuring multiple AP Signal-Mimosa}
\label{Experiment1}
\end{center}
\end{figure} 

The experimental setup consists of a PhytlSigns biosignal amplifier device developed for the purpose of measuring electrical signals in plants, by a Swiss based technology company Vivent Sarl \cite{Phyt}. This device consists of a plant electrophysiology sensor connected to a single board computer (i.e., Raspberry Pi (RPi) which acts as a data logger. We use two different plants for this work i.e., \textit{Mimosa pudica} and \textit{Aloe Vera}. The RPi is connected to the mains power by a grounded 12 Watt power cable. A mini USB cable connects the PhytlSigns device and the RPi. The PhytlSigns device has two sockets labeled as ground and stalk. Two auxiliary cables are used in this experimental setup. The first cable connects the ground socket to the soil of the plant (to act as the ground for the experiment) whereas the second cable connects the stalk socket to the stalk of the plant under experimentation. A conductive gel is used to gently place the pin on the stalk in order to prevent puncturing or damage to the stalk of the plant. The RPi is connected to the PhytlSigns device as well as the local router with an Ethernet cable. The raw data generated by the experiments is stored within the RPi. This data is also transmitted by the RPi to the server. \textcolor{black}{The IP address of the RPi can be typed into a browser to monitor the change in potentials of the plants in real time. We finally use the raw data stored in the RPi to generate the graphs for voltage signals using MATLAB.}

\begin{figure}
\begin{center}
\includegraphics[trim=0cm 0cm 0cm 0cm ,clip=true, width=0.75\columnwidth]{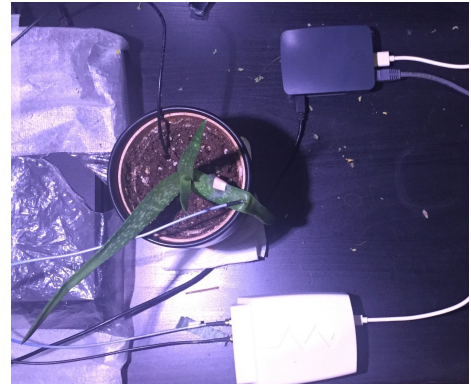}
\caption{Experimental Setup for Measuring Activation Signal- Aloe Vera}
\label{Experiment2}
\end{center}
\end{figure}

\textcolor{black}{In this work, we perform a range of different experiments to measure the amplitudes of the multiple AP signals for \textit{Mimosa pudica} plants as well as multiple activation signals in \textit{Aloe Vera}. Specifically for this paper we conduct three different types of experiments. In the first two cases we apply stimulus with varying intensity resulting in different amplitudes of generated signals for a \textit{Mimosa pudica}. In the third case we repeat the stimulus after a refractive period (the resting period) for both the plants. The aim to these experiments is to verify the theoretical models presented for the generation of multiple AP and multiple activation signals in plants. The raw data generated from all these experiments is collected in the RPi which is then converted into graphs using MATLAB. We also use this experimental data to compute the mutual information and compare it with the results obtained from the numerical method in Section VIII. }

\begin{table}[]
\centering
\caption{Parameters and their default values for AP Signals.}
\begin{tabular}{|c|c|}
\hline
\multicolumn{1}{|c|}{Symbols}	&	\multicolumn{1}{|c|}{Notation and Value}	\\
 \hline
$E_R$ &     Resting Potential  = -150-170 mV
\\ \hline
$F$ &    Faraday's constant = $9.65 \times 10^4 C/mol$ 
\\ \hline
$C$ &     Membrane capacitance = $10^{-6} F cm^-2$  
\\ \hline
$P_m$  &     Permeability per unit area  = $10^{-6}$ M cm $s^{-1}$
\\ \hline
$\gamma$ &  ratio of  rate constants = $9.9 \times 10^-5 M$
\\ \hline
 $\phi_{i}$ & Probability ion link - inside = $c_{in}$ / ($c_{in}$ + $\gamma$)
\\ \hline
 $\phi_{o}$ & Probability ion link - outside =  $c_{out}$ / ($c_{out}$ + $\gamma$)
 \\ \hline
 $\eta_i$ & Probability ion not linked-inside = 1- $\phi_{i}$ 
\\ \hline
 $\eta_o$ & Probability ion not linked-outside = 1- $\phi_{o}$
\\ \hline
$c_{in}$ and $c_{out}$ &  1.28	and 1.15 respectively
\\ \hline
$z$ & ion charge e.g. for calcium = +2.  
\\ \hline
$p_o$ & ion channel open-state probability 
\\ \hline
\end{tabular}
\label{table:1}
\end{table}
\begin{table}[]
\centering
\caption{Parameters and their default values for Mechanosensitive Activation Signal.}
\begin{tabular}{|c|c|}
\hline
\multicolumn{1}{|c|}{Symbols}	&	\multicolumn{1}{|c|}{Notation and Value}	\\
 \hline
$P_m$  &     Permeability per unit area  = $10^{-6}$ M cm $s^{-1}$
\\ 
 \hline
$F$ &    Faraday's constant = $9.65 \times 10^4 C/mol$ 
\\ \hline
$K_B$  &     Boltzmann Constant  = $1.3807 \times 10 ^ -23 J K^-1 $
\\ \hline
 $\phi_{i}$ & Probability ion link - inside = $c_{in}$ / ($c_{in}$ + $\gamma$)
\\ \hline
 $\phi_{o}$ & Probability ion link - outside =  $c_{out}$ / ($c_{out}$ + $\gamma$)
 \\ \hline
 $\eta_i$ & Probability ion not linked-inside = 1- $\phi_{i}$ 
\\ \hline
 $\eta_o$ & Probability ion not linked-outside = 1- $\phi_{o}$
\\ \hline
$z$ & ion charge e.g. for calcium = +2.  
\\ \hline
\end{tabular}
\label{table:2}
\end{table}

\section{ Results and Discussions}
\label{numerical}
 
In this section we present the numerical results for the system model considered in this paper. The parameters used for the generation of multiple AP and multiple activation signal are presented in Tables \ref{table:1} and \ref{table:2} respectively. For the propagation medium we assume a voxel size of ($\frac{1}{3}$ $\mu$m)$^{3}$ (i.e., $\Delta = \frac{1}{3}$ $\mu$m), creating an array of $3\times 3 \times 1$ voxels for different receiver configurations.  We assume one transmitter cell and increasing number of receiver cells in different configurations up to 100. The transmitter and each receiver cell occupy one voxel each as described in the system model in Section \ref{system}. The mean emission rate $c$ is dependent on the AP or activation signal which triggers the release of molecules. The aim is to compute the mutual information between the input and output number of molecules of the system and the information propagation speed associated with different signals.

\textcolor{black}{First in Figures 7 and 8 we calculate the magnitude of a single AP signal from the numerical model and compare it with corresponding experimental calculation (using PhytlSigns device) for two different intensity levels of stimulus in a \textit{Mimosa pudica} plant. This results in different amplitudes of a single AP signal, from 30 to 70 mV. We observe that as the intensity of stimulus increases, the experimental and calculated results become closer. The slightly slower decay of the higher amplitude AP signals can be understood from the mathematics of the AP generation model.} 

\textcolor{black}{Next in Figure 9 we compare the numerical and experimental results for the case of multiple APs in a \textit{Mimosa pudica} plant. We observe that experimental results closely match with the theoretical results.}  Figure \ref{info} shows the comparison of the maximum mutual information calculated from the experimental data and numerically obtained values for a \textit{Mimosa} plant. This validates the mutual information calculated by using the theoretical model.

\begin{figure}
\begin{center}
\includegraphics[trim=0cm 0cm 0cm 0cm ,clip=true, width=0.8\columnwidth]{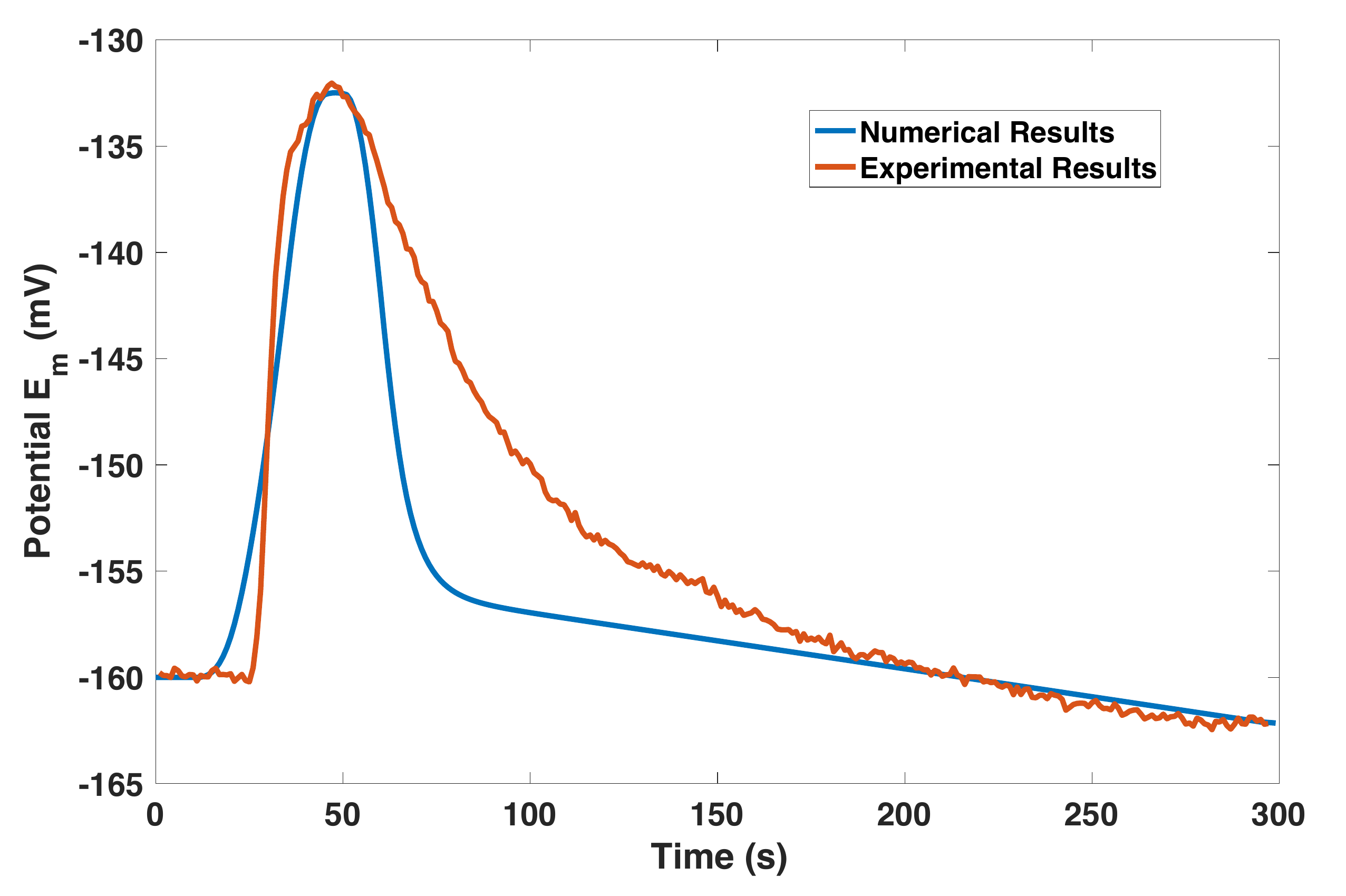}
\caption{Comparison- Moderate Stimulus experimental and numerical results}
\label{compare1}
\end{center}
\end{figure}

\begin{figure}
\begin{center}
\includegraphics[trim=0cm 0cm 0cm 0cm ,clip=true, width=0.8\columnwidth]{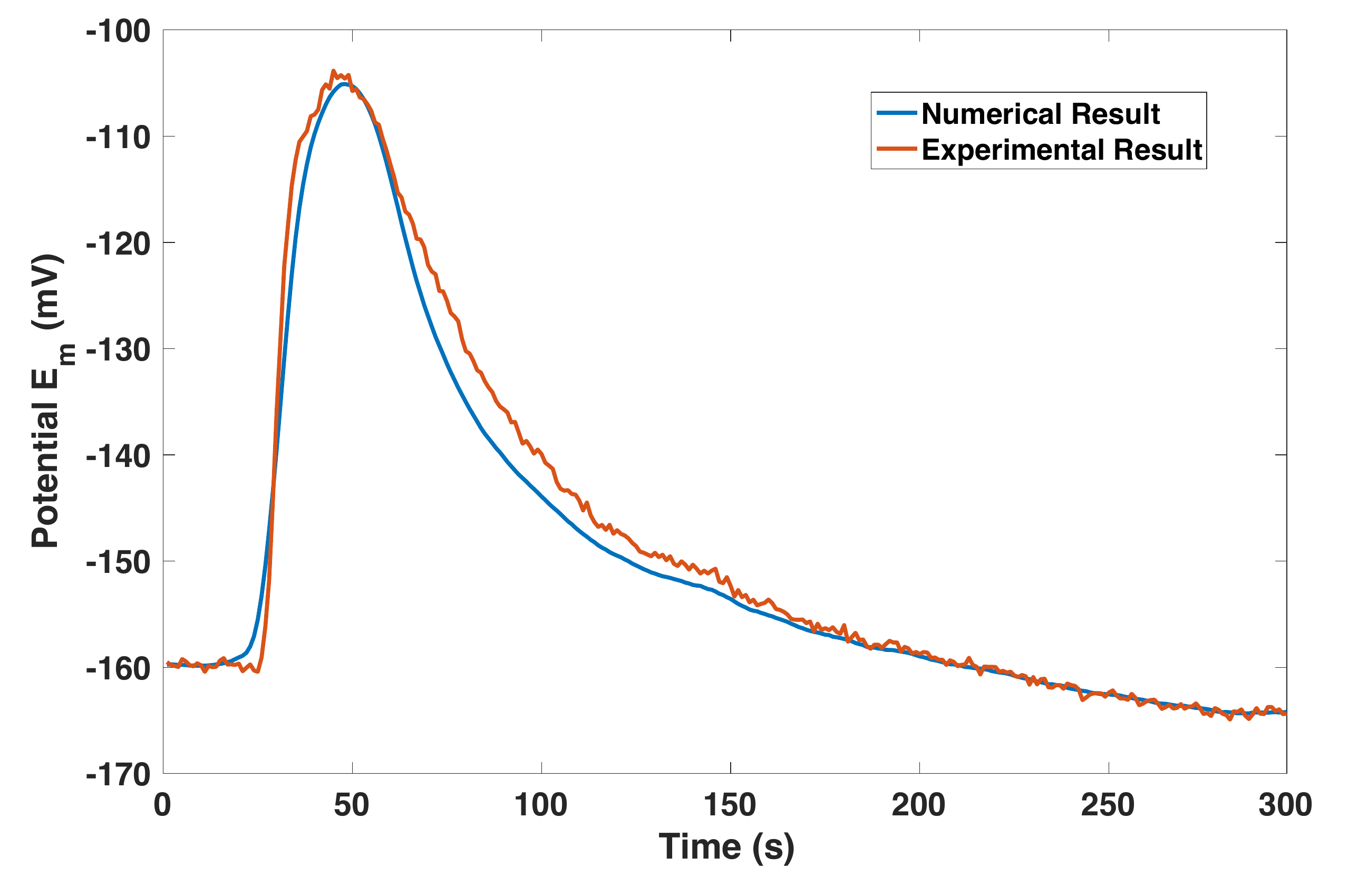}
\caption{Comparison- Intense Stimulus experimental and numerical results}
\label{compare}
\end{center}
\end{figure}

\begin{figure}
\begin{center}
\includegraphics[trim=0cm 0cm 0cm 0cm ,clip=true, width=0.8\columnwidth]{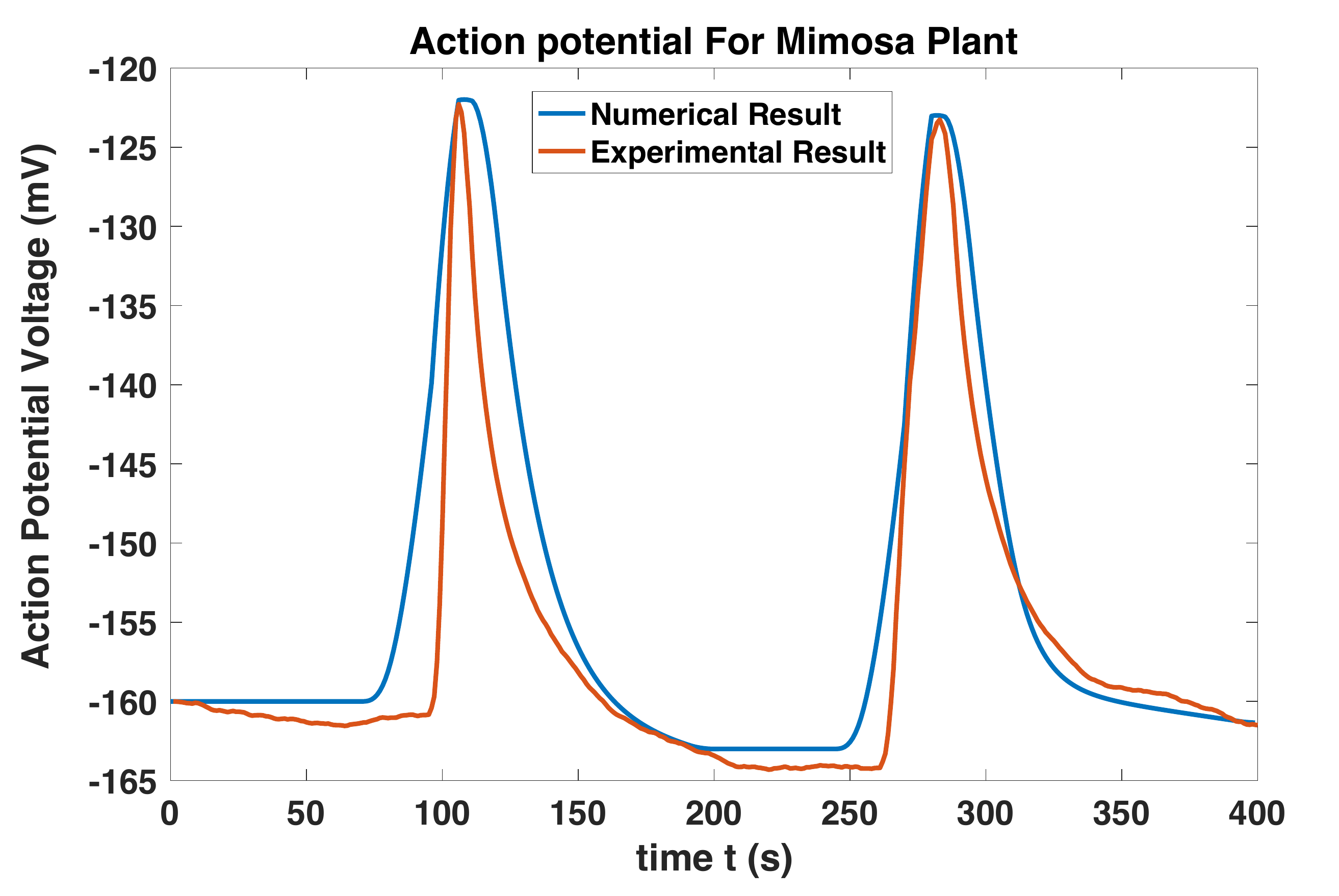}
\caption{Comparison of experimental and numerical results-Multiple AP}
\label{compare2}
\end{center}
\end{figure}

\begin{figure}
\begin{center}
\includegraphics[trim=0cm 0cm 0cm 0cm ,clip=true, width= 0.8\columnwidth]{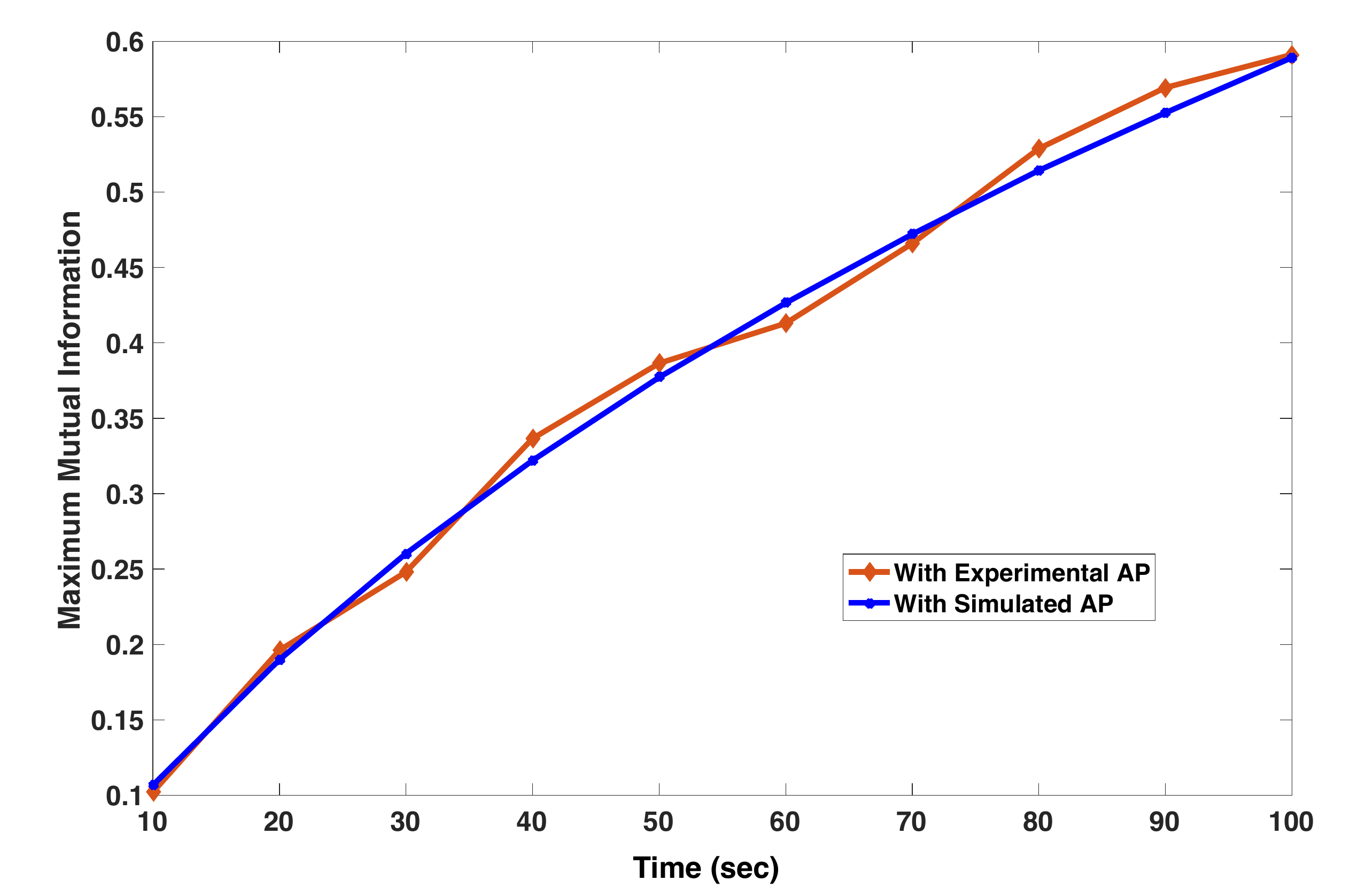}
\caption{Mutual Information comparison for 3 receiver cells in series}
\label{info}
\end{center}
\end{figure}

\begin{figure}
\begin{center}
\includegraphics[trim=0cm 0cm 0cm 0cm ,clip=true, width=0.8\columnwidth]{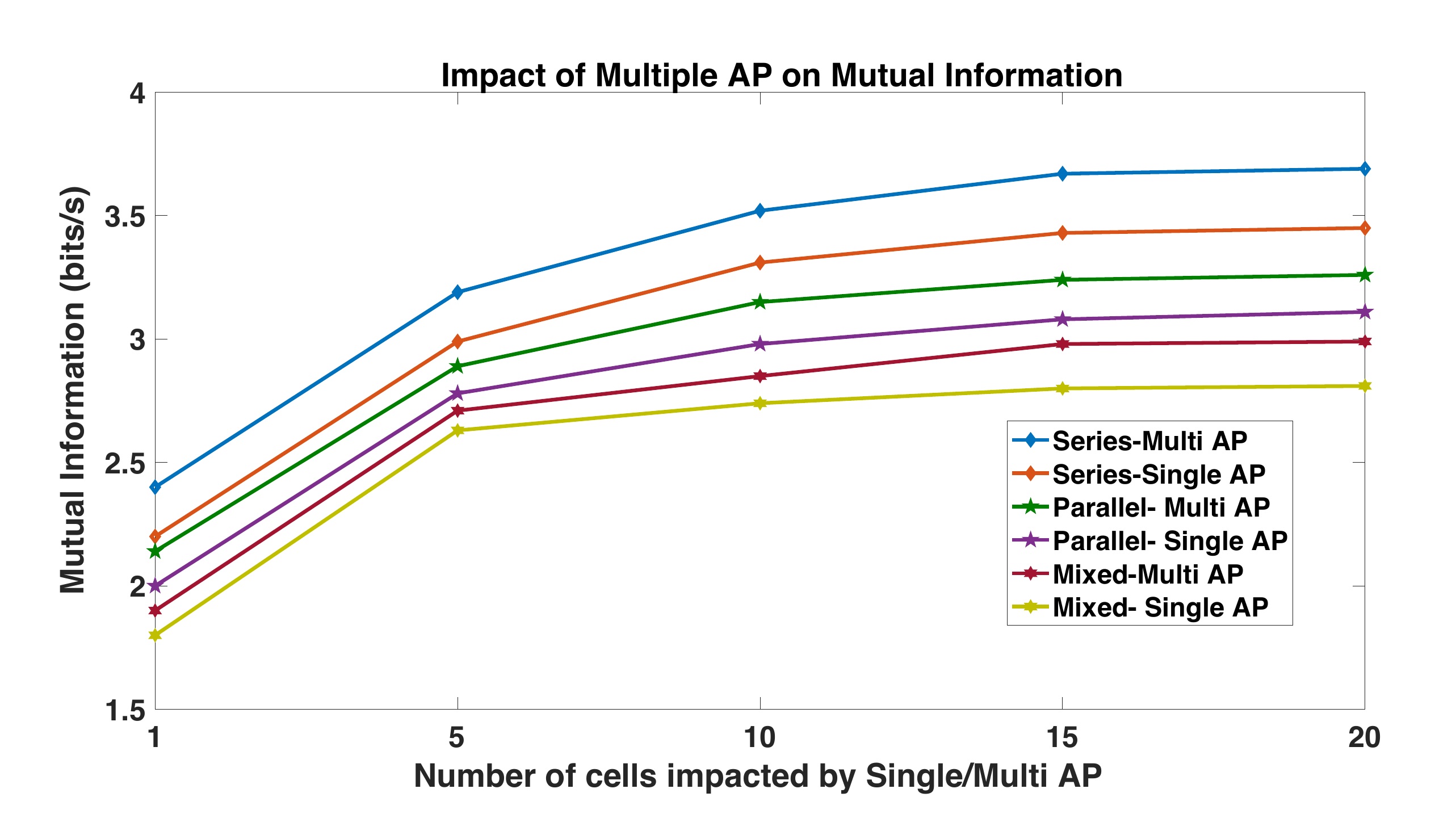}
\caption{Mutual Information for Single/Multiple AP Signals}
\label{Result 1}
\end{center}
\end{figure}

Next in Figures 11 and 12 we study the impact of multiple AP signals on the mutual information and the information propagation speed for increasing number of receiver cells (in different configurations) up to 100. Note that we show the results for only up to 20 cells as the results remain constant for 20-100 cells. We observe that mutual information is minimum for the case of the mixed receiver configuration for both single and multiple AP. In the case of multiple APs the mutual information is higher for all the configurations as compared to a single AP. This indicates the contribution of secondary AP. For multiple AP, the mutual information is highest for the series configuration of receiver cells. The results in Figure 12 show that for multiple AP case, the propagation speed is higher as compared to the case of single AP. We observe that as the number of cells increase beyond 10-12 the mutual information and the information propagation speed become constant. This means that mutual information per cell decreases when the number of cells increase beyond 10-12.

\begin{figure}
\begin{center}
\includegraphics[trim=0cm 0cm 0cm 0cm ,clip=true, width=1\columnwidth]{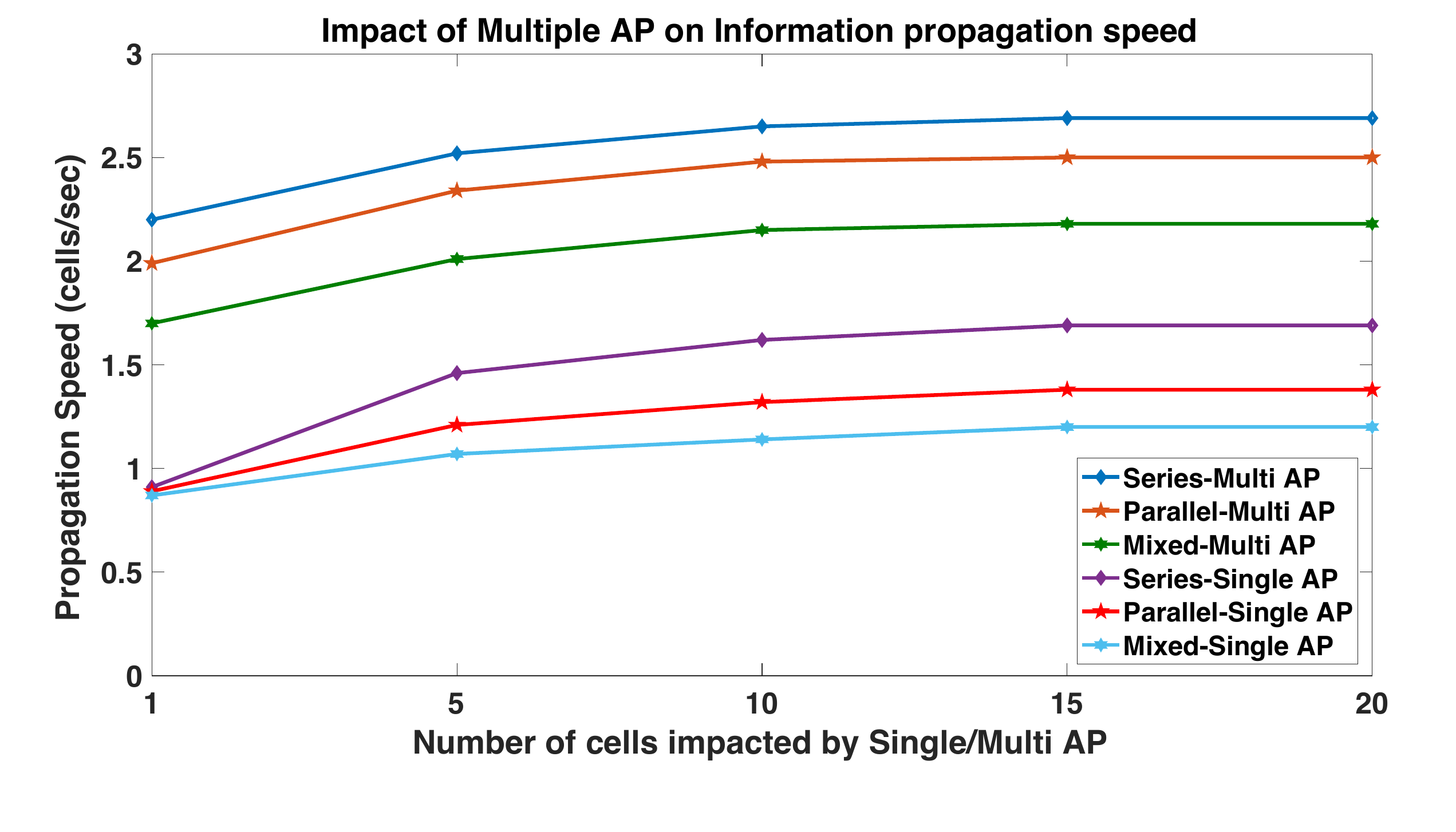}
\caption{ Information Propagation Speed for Single/Multiple AP Signals}
\label{Result 2}
\end{center}
\end{figure}

\begin{figure}
\begin{center}
\includegraphics[trim=0cm 0cm 0cm 0cm ,clip=true, width=0.8\columnwidth]{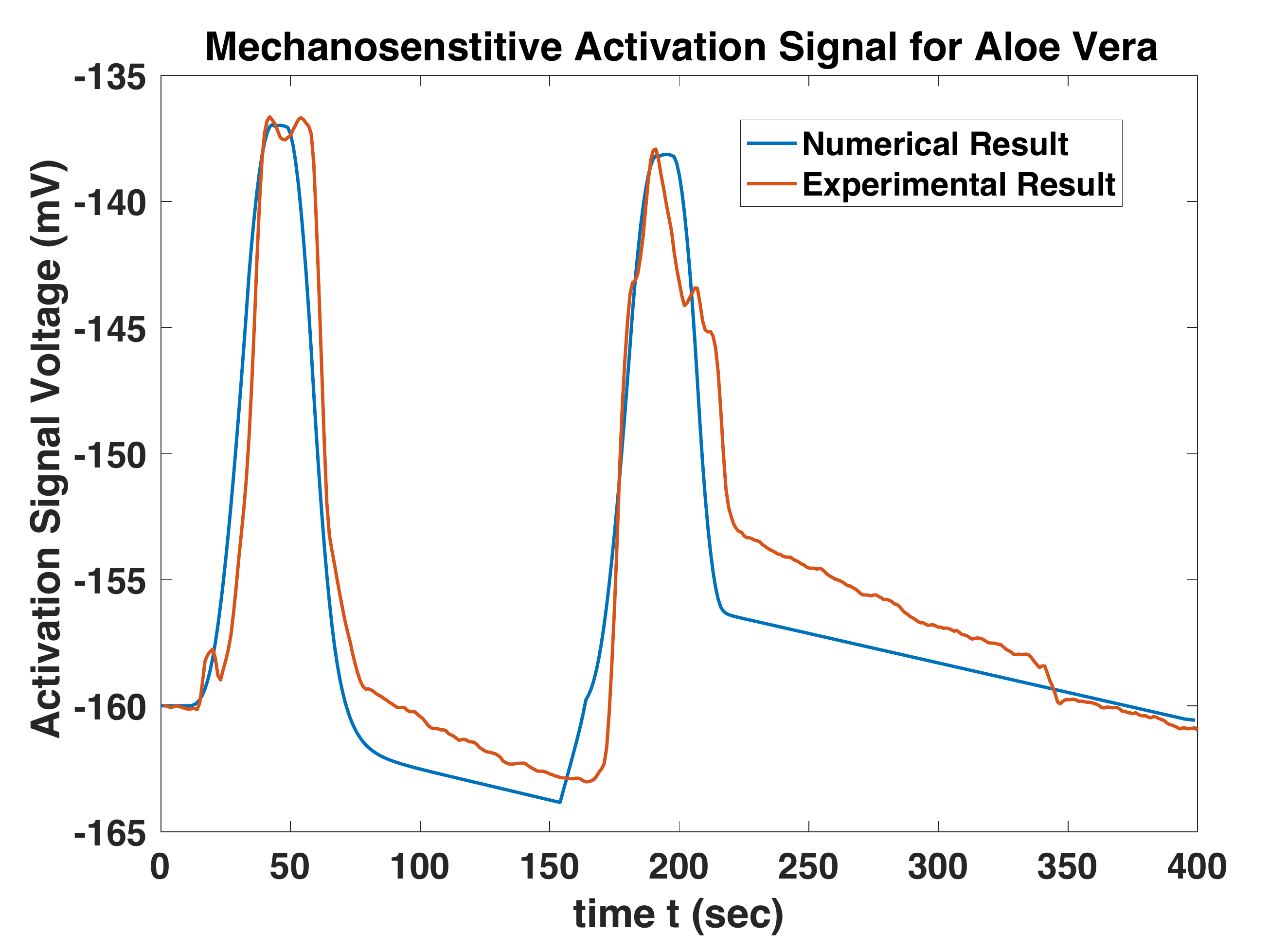}
\caption{Comparison of experimental and numerical results-Multiple Activation Signal}
\label{current}
\end{center}
\end{figure}

Next in Figure 13 we present the experimental and numerical results for the voltage associated with the multiple mechanosensitive activation signals generated by a mechanical stimulus. The experimental results are obtained for an \textit{Aloe Vera} plant by using the Phytlsigns device. We observe that both the experimental and numerically calculated result match closely with each other, therefore validating our theoretical model. We also observe that as the stimulus increases the activation signal voltage corresponding to mechanosensitive activation signals increases fastly. 
 
Finally in Figures 14 and 15 we study the impact of single and multiple mechanosensitive activation signals on the mutual information and the information propagation speed. Although we consider 100 cells we only show results up to 20 as the results remain constant for 20-100. We  observe that both the mutual information and information propagation speed tend to increase in the presence of multiple activation signals in the system up to 10-12 cells. However, as the number of cells increase beyond 10-12 the mutual information and the information propagation speed become constant. This means that similar to the case of multiple AP, the mutual information per cell for multiple activation signals tends to decrease as the number of cells increase beyond 10-12. 

\textcolor{black}{Note that the results in Figures 11 and 14 show the results for \textbf{mutual information only}. It can be understood from both these graphs that mutual information per cell (i.e., mutual information/no. of cells) tends to increase initially. However, as the number of cells increase beyond 10-12 the mutual information per cell starts to decrease as the mutual information becomes constant for increasing number of cells. Note that we have obtained similar pattern of results for all three configurations of receiver cells considered in this paper.}

\textcolor{black}{To summarize, the results in Figures 7-9 show the single and multiple action potential signals (both numerical and experimental results) in Mimosa pudica plant  whereas the simulation results in Figure 13 show the mechanosensitive signals (both numerical and experimental results) in an Aloe Vera plant. Whereas the results in Figures 10-12 and 14-15 are based on the numerical calculations only. }

\begin{figure}
\begin{center}
\includegraphics[trim=0cm 0cm 0cm 0cm ,clip=true, width=1\columnwidth]{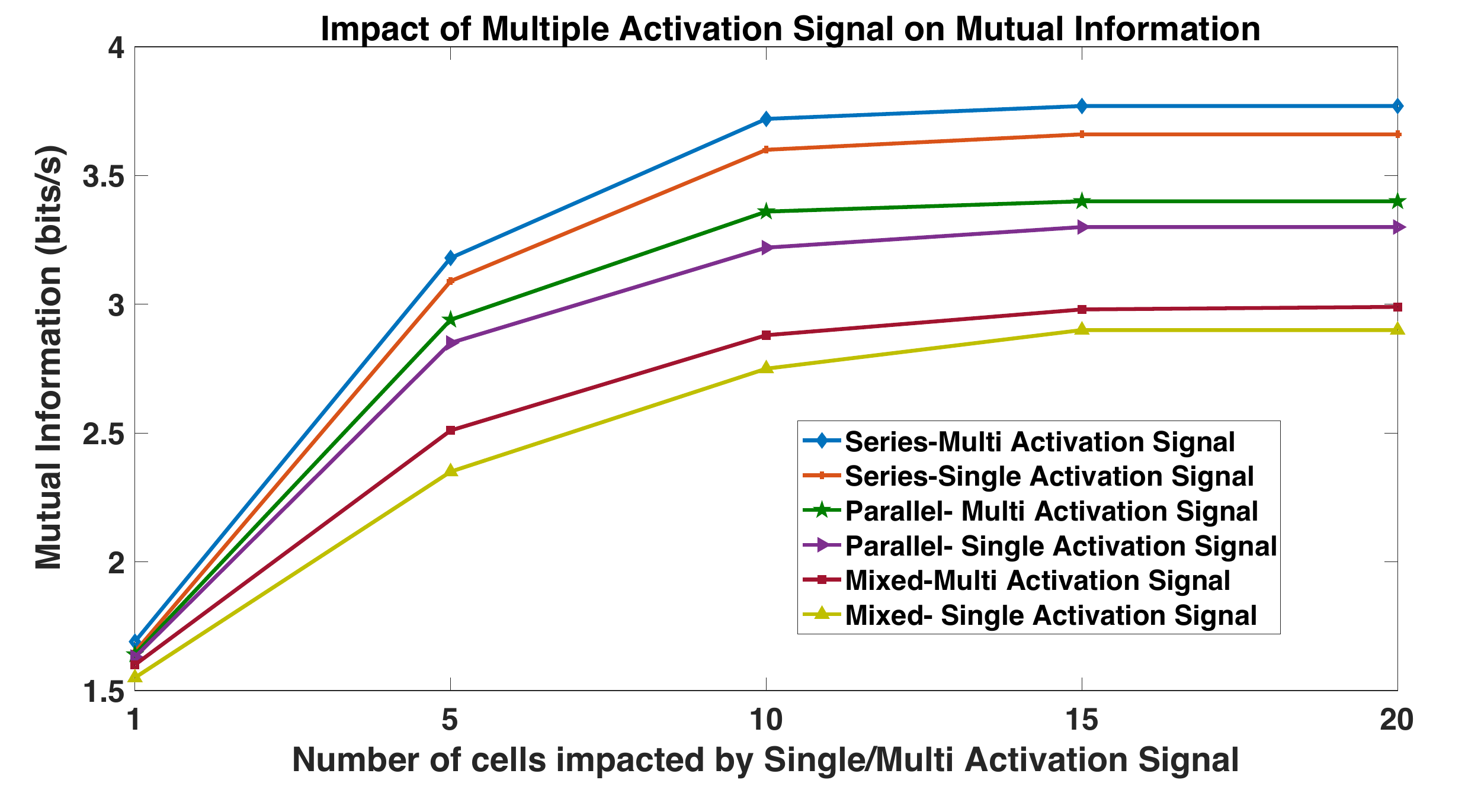}
\caption{Mutual Information for Single/Multiple Activation Signals}
\label{Result 3}
\end{center}
\end{figure}

\begin{figure}
\begin{center}
\includegraphics[trim=0cm 0cm 0cm 0cm ,clip=true, width=1\columnwidth]{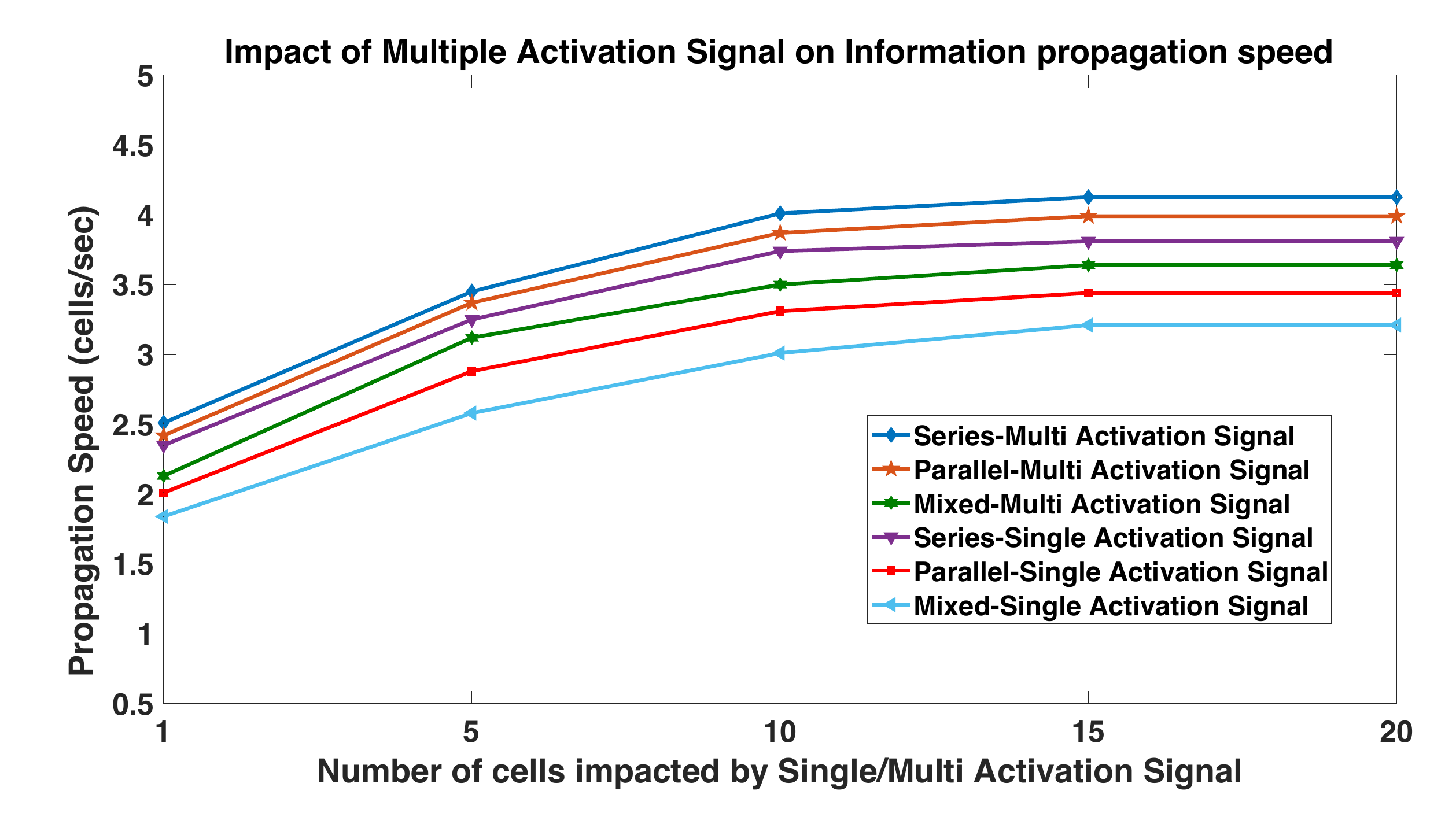}
\caption{Information Propagation Speed for Single/Multiple Activation Signals}
\label{Result 1c}
\end{center}
\end{figure}

\section{Conclusion}
\label{conclusion}
In this paper we study the impact of multiple APs and multiple mechanosensitive activation signals on the inter-cellular communication in plants. We show that when multiple APs or multiple activation signals are generated due to an external stimulus,  there is a general increasing trend in the mutual information of the system with an increase in the population of cells up to certain length. Moreover we show that the information propagation speed also increases in the presence of the multiple AP or activation signals in the system with an increase in the population of cells up to certain length. We also verify the results produced by the theoretical model with experiments on different plants such as \textit{Mimosa pudica} and \textit{Aloe Vera}. The experimental and numerical results, on the signal voltage, mutual information and information propagation speed, are important contributions to information theoretic modelling of biological communication. Especially these results will help the researchers to investigate and develop applications related to inter-cellular communication signals in biological systems like plant cell networks. Furthermore the validation of numerical results by experimentation confirms the relevance of information theoretic modeling as a tool to better understand real plant signals and their applications.

\ifCLASSOPTIONcaptionsoff
  \newpage
\fi

\bibliographystyle{ieeetr}
\bibliography{nano2017,book,nano2018,1,2,3,4,5,6}

\end{document}